\documentclass{iaus}
\usepackage{graphicx}

\title[Metals in High-$z$ Galaxies] 
{Metals in Star-Forming Galaxies at High Redshift}

\author[Claus Leitherer]   
{Claus Leitherer}

\affiliation{Space Telescope Science Institute, 3700 San Martin Dr., Baltimore, MD 21218, USA
 \break email: leitherer@stsci.edu\\[\affilskip]}

\pubyear{2005}
\volume{228}  
\pagerange{1--7}
\date{?? and in revised form ??}
\jname{From Lithium to Uranium: Elemental Tracers of Early Cosmic Evolution}
\editors{V. Hill, P. Fran\c{c}ois \& F. Primas, eds.}
\begin{document}

\maketitle

\begin{abstract}
The chemical composition of high-redshift galaxies is an important property that gives clues to their past history and future evolution. Measuring abundances in distant galaxies with current techniques is often a challenge, and the canonical metallicity indicators can often not be applied. I discuss currently available metallicity indicators based on stellar and interstellar absorption and emission lines, and assess their limitations and systematic uncertainties. Recent studies suggest that star-forming galaxies at redshift around 3 have heavy-element abundances already close to solar, in agreement with predictions from cosmological models.
\end{abstract}

\firstsection 
\section{Introduction}

Until recently, abundance determinations at cosmological redshifts ($z \gtrsim 2$) were limited to bright quasars and the material along their sightlines. Over the past 5~--~10 years, the number and nature of accessible objects has expanded dramatically, and ``normal'' star-forming galaxies at high redshift have become targets of abundance studies (Pettini 2004). This progress was made possible by the use of efficient search techniques for finding large numbers of candidate galaxies (Steidel et al. 1999) and by the availability of 8~--~10-m class telescopes for spectroscopic studies. I will discuss the general properties of these galaxies, introduce the techniques used for determining abundances, and provide an overview of their chemical composition. I will conclude with a cosmological perspective.

\section{Sample}

The galaxies discussed in this paper are commonly referred to as {\em Lyman-break galaxies} (LBG; Giavalisco 2002) because of the search technique employed to detect them. They are color-selected via their strong Lyman discontinuity and/or their blue restframe ultraviolet (UV) colors. These galaxies are sufficiently bright for obtaining spectroscopic data ($25^m \gtrsim R \gtrsim 21^m$) --- a necessary prerequisite for abundance analyses. Owing to the selection criterion, they are relatively dust-poor with typical visual attenuations of $A_{\rm V} \simeq 1^m$ and are fairly luminous, but not ultraluminous ($L \simeq 10^{11}$~L$_\odot$). LBGs are actively star-forming at rates of order $10^2$~M$_\odot$~yr$^{-1}$ (Pettini et al. 2001) and have correspondingly short recycling times. Therefore the chemical composition of the newly formed stars and the interstellar gas are identical.

Most LBGs are in the redshift range $2 \lesssim z \lesssim 3.5$. The lower and upper limits are imposed by atmospheric cut-offs in ground-based observations and decreasing efficiencies of the search techniques. The corresponding lookback times in a standard universe are 11~Gyr~$\lesssim t \lesssim$~12.5~Gyr, or about 15\% of the age of the universe after its formation. 

\begin{figure}
\begin{center}
 \includegraphics[angle=90,width=0.75\textwidth]{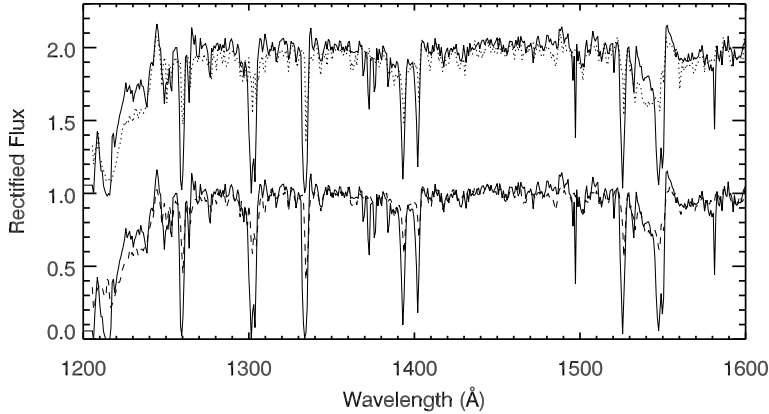}
\end{center}
  \caption{Comparison between the observed spectrum of MS1512--cB58 (solid) and two synthetic models 
with 1/4~Z$_\odot$ (lower; dashed) and Z$_\odot$ (upper; dotted). 
The models have continuous star formation, age 100 Myr, and Salpeter IMF between 1 and 100 M$_\odot$. The stellar lines
are weaker in the metal-poor model (from Leitherer et al. 2001).}
\end{figure}

\section{Techniques --- Restframe Optical versus UV}

Abundance determinations typically fall into two categories, either relying on indicators in the restframe optical, or on those in the restframe UV. The restframe optical wavelength region has traditionally been used to determine galaxy abundances from nebular emission lines. At a redshift of $z = 3$, the restframe optical is observed in the near-infrared (IR) H and K bands. Spectroscopic observations of LBGs in the near-IR have become technically feasible (e.g., Pettini et al. 2001) but abundance analyses are still challenging. Only the strongest lines such as, e.g., H$\alpha$, H$\beta$, [N~II] $\lambda$6584, or [O~III] $\lambda$5007 are detectable at sufficient S/N. Even when good-quality spectra are available, the atmospheric windows usually restrict the wavelengths to a narrow range, which precludes commonly used techniques such as the classical R23 strong-line method (McGaugh 1991). 

The need for alternative variants of the classical strong-line method led Pettini \& Pagel (2004) to readdress the usefulness of the N2 and O3N2 ratios. The former is defined as the ratio [N~II] $\lambda$6584 over H$\alpha$ and was recently discussed by Denicol\'o et al. (2002); the latter includes the oxygen line for the ratio ([O~III]~$\lambda$5007/H$\beta$)/([N~II]~$\lambda$6584/H$\alpha$) and was originally introduced by Alloin et al. (1979). After calibrating the two abundance indicators with a local H~II region sample, Pettini \& Pagel find that O3N2 and N2 predict O/H to within 0.25~dex and 0.4~dex at the 2~$\sigma$ confidence level, respectively.

The observed frame optical wavelength region corresponds to the restframe UV of LBGs. The UV contains few nebular emission lines in star-forming galaxies (Leitherer 1997) and has rarely been used for chemical composition studies in {\em local} galaxies of this type. Fig.~1 compares the UV spectrum of the LBG MS1512--cB58 with theoretical spectra (Leitherer et al. 2001). Three groups of lines can be distinguished: (i) Interstellar absorption lines, most of which are strong and heavily saturated. Only in very few cases can unsaturated absorption lines in LBGs be used for an abundance analysis. (ii) Broad stellar-wind lines with emission and blueshifted absorption. These lines are the telltales of massive OB stars whose stellar winds are metallicity dependent. (iii) Weak photospheric absorption lines which can only be seen in high-quality spectra. 
Abundance studies from stellar lines in restframe UV spectra must rely either on suitable template stars or on extensive non-LTE radiation-hydrodynamic models which are only beginning to become available (Rix et al. 2004).

\section{The Chemical Composition of LBGs}

An initial, rough estimate of the heavy-element abundances can be obtained from the equivalent widths of the strong UV absorption lines. Heckman et al. (1998) pointed out the close correlation of the Si~IV $\lambda$1400 and C~IV $\lambda$1550 equivalent widths with O/H in a sample of {\em local} star-forming galaxies. This correlation seems surprising, as these stellar-wind lines are deeply saturated. The reason for the metallicity dependence is the behavior of stellar winds in different chemical environments. At lower abundance, the winds are weaker and have lower velocity, and the lines become weaker and narrower. As a result, the equivalent widths are smaller at lower O/H. If the same correlation holds at high redshift, the observed equivalent widths in LBGs suggest [O/H]~$\simeq$~--0.5 (Leitherer 1999). A similar, somewhat weaker correlation exists between O/H and the equivalent widths of the strongest {\em interstellar} lines. This is even more unexpected because the equivalent widths of saturated lines have essentially no dependence on the column density $N_{\rm ion}$: $W \propto b[\ln(N_{\rm ion}/b)]^{0.5}$. Therefore the correlation must be caused by the $b$ factor, and therefore by velocity. More metal-rich galaxies are thought to host more powerful starbursts with correspondingly larger mechanical energy release by stellar winds and supernovae. The energy input leads to increased macroscopic turbulence and higher gas velocities at higher O/H (Heckman et al. 1998). If the same applies to star-forming galaxies in the high-redshift universe, their measured equivalent widths again indicate an oxygen abundance of about 1/3 the solar value. 

Pettini et al. (2001) determined oxygen abundances in five LBGs from emission lines in restframe optical spectra. The redshift range of the sample dictated the use of the R23 method. The galaxies turned out to be rather metal-rich, with O/H somewhat below the solar value. This is roughly in agreement with restframe UV results, and an order of magnitude above the metallicities found in damped Lyman-$\alpha$ absorbers (DLA) which are found at the same redshift. Because of the double-valued nature of the R23 method, the possibility exists but is deemed less likely that the sample has oxygen abundances of only 1/10 the solar value.

The lensed LBG MS1512--cB58 and its bright restframe UV spectrum can be studied at sufficiently high S/N and resolution to detect and resolve faint, unsaturated interstellar absorption lines. Pettini et al. (2002) measured numerous transitions from H to Zn covering several ionization stages. Abundances of several key elements could be derived. The $\alpha$-elements O, Mg, Si, P, and S all have abundances of about 40\% solar, indicating that the interstellar medium  is highly enriched in the chemical elements produced by type II supernovae. In contrast, N and the Fe-peak elements Mn, Fe, and Ni are all less abundant than expected by factors of several. In standard chemical evolution models, most of the nitrogen is produced by intermediate-mass stars, whereas type Ia supernovae contribute most of the Fe-peak elements. Since the evolutionary time scales of intermediate- and low-mass stars are significantly longer than those of massive stars producing the $\alpha$-elements, the release of N and the Fe-group elements into the interstellar medium is delayed by $\sim$10$^9$~yr. MS1512--cB58 may be an example of a star-forming galaxy in its early stage of chemical enrichment, consistent with its cosmological age of only about 15\% of the age of the universe. Mehlert et al. (2002) provided similar arguments to explain variation of the  C~IV $\lambda$1550 line relative to Si~IV $\lambda$1400 in a small sample of LBGs. C~IV appears to decrease in strength relative to Si~IV from lower to higher redshift, which may reflect the time delay of the carbon release by intermediate-mass stars.

\begin{figure}
\includegraphics[width=0.5\textwidth]{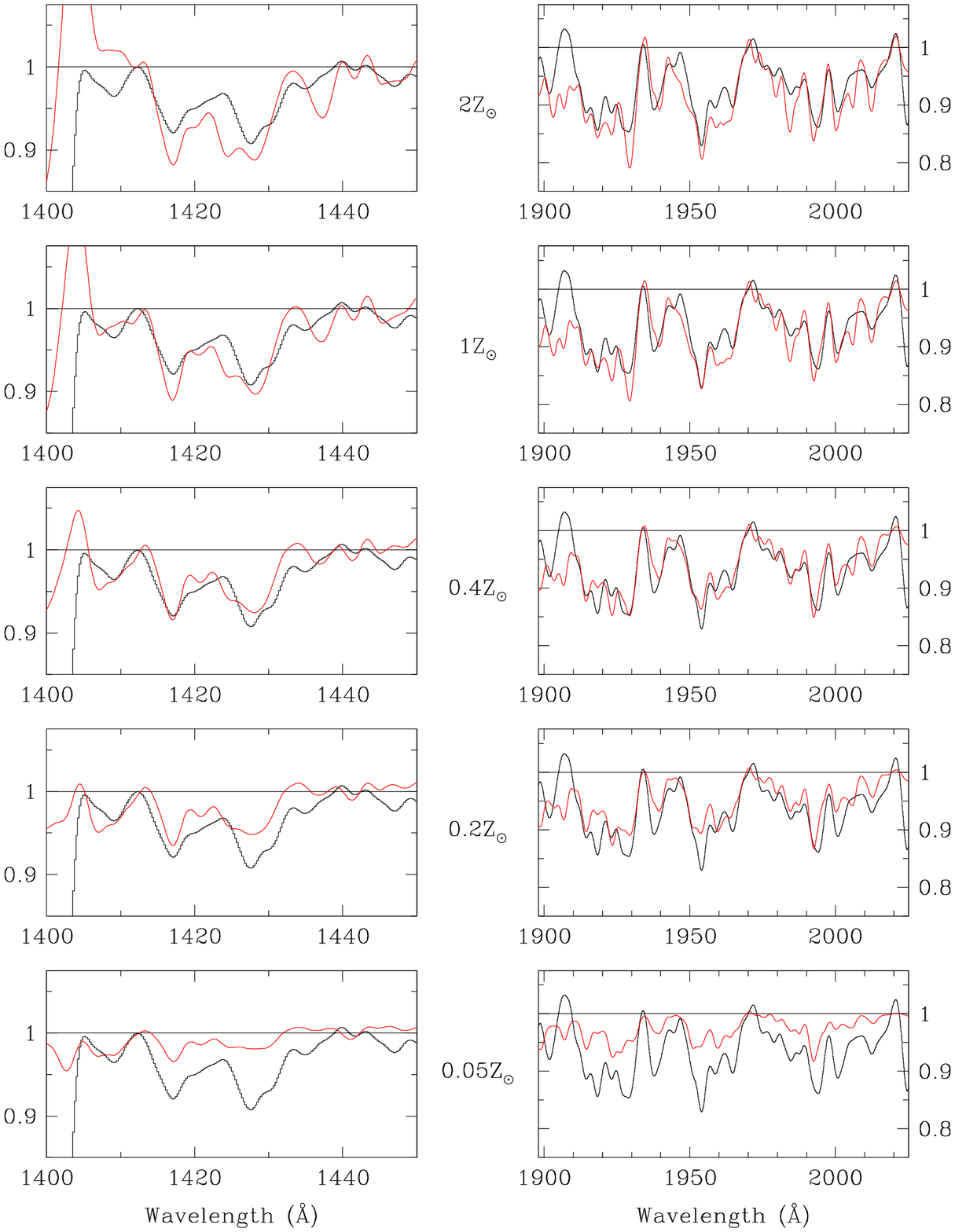}\includegraphics[width=0.5\textwidth]{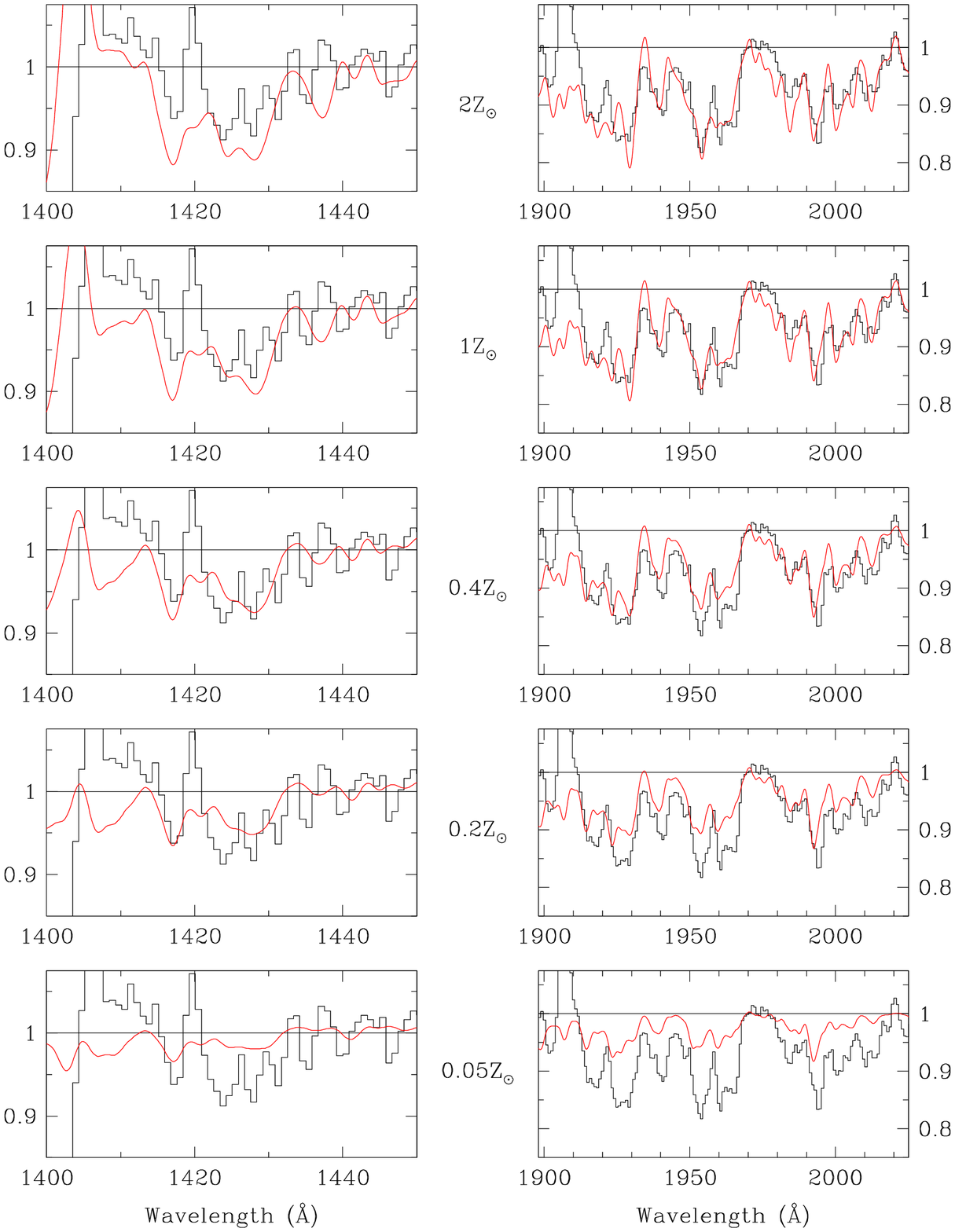}
  \caption{Left pair of panels: comparison of the observed spectrum of MS1512--cB58 (thick) with fully synthetic spectra (thin) for five different metallicities, from twice solar to 1/20 solar. First panel: region around 1425~\AA; second: region of the Fe III blend near 1978~\AA. Each pair of panels is labeled with the metallicity of the synthetic spectrum shown. Right pair of panels: same as left pair, but for Q1307--BM1163 (from Rix et al. 2004).}
\end{figure}

The interstellar lines in LBGs have blueshifts with velocities of up to several hundred km~s$^{-1}$ indicating large-scale outflows. The associated galactic mass-loss rates of $\sim$10$^2$~M$_\odot$~yr$^{-1}$ are comparable to the rates of star formation. The newly formed heavy elements are removed from their birth sites by stellar winds and supernovae and are
transported into the halo and possibly into the intergalactic medium (Pettini et al. 2002).

Detailed studies of weak interstellar lines such as that done for MS1512--cB58 remain a technical challenge, even for high-throughput spectrographs at the largest telescopes. Furthermore, the results for Fe-peak elements carry some uncertainty because of the a priori unknown depletion corrections. Abundance analyses using stellar lines are not affected by depletion uncertainties. However, the existence of non-standard element ratios precludes the use of locally observed template spectra for spectral synthesis. Therefore our group (F. Bresolin, R. Kudritzki, C. Leitherer, M. Pettini, S. Rix) has embarked on a project to model the spectra of hot stars and link them with a spectral synthesis code to predict the emergent UV spectrum of a composite stellar population as a function of metallicity. We generated a grid of hydrodynamic non-LTE atmospheres with the WM-basic code (Pauldrach et al. 2001) and calculated the corresponding UV line spectra. The resulting library was incorporated into the Starburst99 code (Leitherer et al. 1999) which then allowed us to compute a suite of model spectra for appropriate stellar population parameters. As a first application, we used several faint stellar blends around 1425~\AA\ and 1978~\AA\ as a metallicity indicator (Rix et al. 2004). The 1425~\AA\ feature is a blend of Si~III, C~III, and Fe~V, and the 1978~\AA\ absorption is mainly Fe~III. The synthesized spectra for five metallicities are compared to the observed restframe UV spectra of MS1512--cB58 and Q1307--BM1163 in Fig.~2. The model having 40\% solar metal abundance provides the best fit to the data, in agreement with the results from other methods.

A variety of independent techniques lead to consistent results for the chemical composition of LBGs. While each method by itself is subject to non-negligible uncertainties, the overall agreement of the results gives confidence in the derived abundances. LBGs at $z \simeq 3$ have heavy-element abundances of about 1/3 the solar value.

\section{Cosmological Perspective}

\begin{figure}
\begin{center}
 \includegraphics[width=0.65\textwidth]{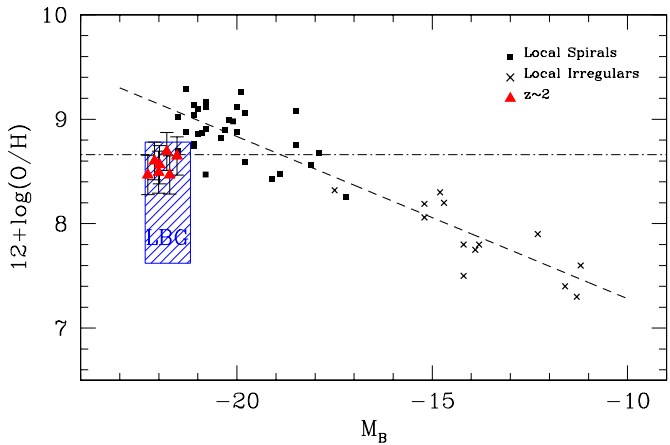}
\end{center}
  \caption{Metallicity-luminosity relationship. Data for {\em local} spiral and irregular galaxies are from Garnett (2002). The $z = 2$ objects are overluminous for their (O/H) abundances, derived using the N2 calibration of Pettini \& Pagel (2004) but lie closer to the relationship for the local galaxies than $z=3$ LBGs (from Shapley et al. 2004).}
\end{figure}

Star-forming galaxies at $z \simeq 3$, at an epoch when the universe's age was only 15\% the present value, display a high level of chemical enrichment. What does their chemical composition tell us about their relation to other galaxies at lower redshift and to other structures found at $z = 3$?

Galaxies at somewhat lower redshift have only recently become accessible for detailed study due to the combined challenges of instrumentation and the galactic spectral properties. Shapley et al. (2004) obtained K-band spectroscopy of seven UV-selected star-forming galaxies at redshifts between 2 and 2.5. The N2 method calibrated by Pettini \& Pagel (2004) was used as an abundance diagnostic. When compared to the original higher-$z$ LBGs, the $z \simeq 2$ sample is more metal-rich. This can be seen in Fig.~3, where O/H of the $z = 2$ galaxies is compared with that of LBGs at $z \simeq 3$ and of local star-forming galaxies over a range of blue luminosities. The latter were analyzed with the R23 method. The $z = 2$ sample has almost solar chemical composition but is still less metal-rich than local late-type galaxies {\em with comparable luminosities}. As a caveat, the comparison rests on the assumption that the N2 and R23 calibrations have no significant offset. The difference between the average redshift of the LBG sample and of the $z = 2$ galaxies translates into a mean age difference of about 1~Gyr. Both the chemical properties and the masses of the $z = 2$ galaxies and LBGs are consistent with standard passive evolution models. 

Kewley \& Kobulnicky (2005) followed the metallicity evolution of star-forming galaxies with comparable luminosities from $z = 0$ to 3.5. O/H was determined from restframe optical emission lines using the strong-line method in four homogeneous galaxy samples. The samples were taken from the CfA2 survey, from the GOODS field, from Shapley et al. (2004), and from the LBG sample, covering $z \approx 0$, 0.7, $2.1 - 2.5$, and $2.5 - 3.5$, respectively. The average oxygen abundance in the local universe, as defined by the CfA2 sample is about solar. O/H decreases with redshift to approximately 1/3 solar at $z=3$.

It is instructive to compare the heavy-element abundances of LBGs to those of DLAs and to the Lyman-forest at the same redshift (Pettini 2004). DLA systems have metallicities of about 1/15~Z$_\odot$ and are thought to be the cross sections of the outer regions and halos of (proto)-galaxies seen along the sightlines of quasars. Although the properties of LBGs and DLAs do not immediately support a close relation between the two classes of objects, at least some link seems likely. If so, the observed outflows in LBGs may provide the metal enrichment of the halos.
The Lyman-forest is predicted by cold dark matter models to result from structure formation in the
presence of an ionizing background. The Lyman-forest had long been
thought to be truly primordial, but metal enrichment of 1/100~--1/1000~Z$_\odot$ has recently been detected (Aguirre et al. 2004).
This relatively high metal abundance early in the evolution of the universe could have been produced by a first generation of Population III stars. Such stars can account for
the amount of metals, and at the same time could have provided copious ionizing photons, as metal and photon
production are closely correlated. Alternatively, star-forming galaxies at high redshift could be the
production sites of the metals seen in the intergalactic medium {\em if} superwinds are capable of removing the newly formed metals from galactic disks.

\begin{acknowledgments}
I would like to thank Max Pettini for a careful reading of the manuscript and for providing invaluable comments. 
\end{acknowledgments}

\end{document}